\documentclass[aps,prx,twocolumn,final,letterpaper]{revtex4}

\usepackage{appendix}
\usepackage{graphicx}   
\usepackage{import}                         
\usepackage{epstopdf}
\usepackage{amsmath} 
\usepackage{bm}
\usepackage{amssymb}
\usepackage{quotes}
\usepackage{transparent}
\usepackage{dcolumn}
\usepackage{multirow}
\usepackage{cancel} 
\usepackage{mdframed}
\usepackage{color}
\usepackage{bm}
\usepackage{dsfont}
\usepackage{slashed}
\usepackage{enumitem}

\begin{document}
\rmfamily

\title{Deterministic quantum state generators and stabilizers from nonlinear photonic filter cavities}

\author{Sean Chen$^{1,4}$, Nicholas Rivera$^{1,2}$, Jamison Sloan$^{3,4}$, and Marin Solja\v{c}i\'{c}$^{1,3}$}

\address{1. Department of Physics, Massachusetts Institute of Technology, Cambridge MA 02139, USA\\
2. Department of Physics, Harvard University, Cambridge MA 02138, USA\\
3. Research Laboratory of Electronics, Massachusetts Institute of Technology, Cambridge MA 02139, USA\\
4. Department of Electrical Engineering and Computer Science, Massachusetts Institute of Technology, Cambridge MA 02139, USA
}

\renewcommand{\abstractname}{} 
\begin{abstract}
Quantum states of light, particularly at optical frequencies, are considered necessary to realize a host of important quantum technologies and applications, spanning Heisenberg-limited metrology, continuous-variable quantum computing, and quantum communications. Nevertheless, a wide variety of important quantum light states are currently challenging to deterministically generate at optical frequencies. In part, this is due to a relatively small number of schemes that prepare target quantum states given nonlinear interactions. Here, we present an especially simple concept for deterministically generating and stabilizing important quantum states of light, using only simple third-order optical nonlinearities and engineered dissipation. We show how by considering either a nonlinear cavity with frequency-dependent outcoupling, or a chain of nonlinear waveguides, one can ``filter'' out all but a periodic ladder of photon number components of a density matrix. As examples of this phenomenon, we show cavities which can stabilize squeezed states, as well as produce ``photon-number-comb'' states. Moreover, in these types of filter cavities, Glauber coherent states will deterministically evolve into Schrodinger cat states of a desired order. We discuss potential realizations in quantum nonlinear optics. More broadly, we expect that combining the techniques introduced here with additional ``phase-sensitive'' nonlinearities (such as second-order nonlinearity) should enable passive stabilization and generation of a wider variety of states than shown here.
\end{abstract}

\maketitle

Generating light states with nonclassical properties at optical frequencies remains a major goal of quantum science and engineering. Part of the interest lies in the promising applications of non-Gaussian states with no classical analogs (e.g. Fock, Schr\"{o}dinger cat, and Gottesman-Kitaev-Preskill (GKP) states). They have important applications in areas such as Heisenberg-limited metrology, photonic quantum computing, and bosonic codes \cite{gottesman2001encoding,giovannetti2011advances,degen2017quantum,wang2020efficient}. Many of these states can be produced at microwave frequencies by exploiting superconducting circuit platforms (e.g., \cite{hofheinz2008generation,vlastakis2013deterministically,grimm2020stabilization,campagne2020quantum,deng2023heisenberg}). Nevertheless, the deterministic realization of such states at \emph{optical frequencies} is difficult, in part due to a paucity of systems which support few-photon nonlinearities needed to implement efficient non-Gaussian operations. That said, systems with few-photon nonlinearity are now emerging or within reach, including: Rydberg-EIT \cite{peyronel2012quantum}, cavity quantum electrodynamics \cite{walther2006cavity}, and exciton-polariton platforms \cite{delteil2019towards,munoz2019emergence, liran2023electrically}. Given these opportunities, we aim to develop new deterministic schemes to exploit few-photon optical nonlinearities for non-Gaussian state generation. 

\begin{figure}[t]
  \centering
  \includegraphics[width=3.25in]{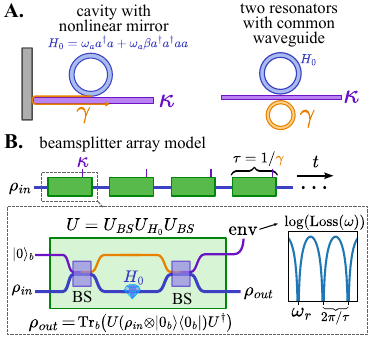}
\caption{\textbf{A.} Examples of ways to realize engineered dissipation: (Left) A nonlinear Kerr cavity (blue) coupled to a waveguide (purple) terminated by a perfect reflector, and (right) two resonators (blue, orange) coupled to a common waveguide. The orange paths have a character frequency $\gamma$ and the waveguide loss occurs with rate $\kappa$. Photons from the Kerr cavity can destructively interfere with photons from orange paths to eliminate loss. 
\textbf{B.} Discrete-time model of nonlinear loss equivalent to continuous nonlinear dissipation, comprised of a sequence of MZIs. Each MZI unit is comprised of a beamsplitter, Kerr nonlinearity, and beamsplitter sequence. For certain frequencies, the loss to the environment goes to zero.}
\label{fig1}
\end{figure}

Here, we introduce a set of schemes for stabilizing a wide range of important quantum states such as Fock, squeezed, and approximate GKP states, based on a combination of engineered dissipation and Kerr (third-order) nonlinearity in the single-photon nonlinear regime. Our technique also deterministically transforms ``classical'' input states (such as coherent states) into nonclassical states, such as $k$-legged cat states, with high fidelity. This scheme, which makes use of engineered dissipation \cite{harrington2022engineered}, stabilizes and generates states without reliance on slow and complex pulse sequences, and may thus enable new opportunities to realize non-Gaussian states with high fidelity and with very high bandwidths.

We start by describing physical systems that implement the engineered dissipation at the heart of our work. One example of such a system is shown in Fig \ref{fig1}A: a nonlinear cavity with frequency-dependent transmission into a waveguide. This can be realized in several ways, e.g. by a resonator (with Kerr nonlinearity) coupled to a waveguide terminated by a mirror, or two resonators (one nonlinear and one linear) coupled to the same waveguide. In either system, if the nonlinear resonator has a certain frequency, it will not dissipate into the waveguide due to interference. Such states are known as bound states in the continuum (BICs), or embedded eigenvalues, because they have in principle infinite lifetime, despite overlapping in frequency with delocalized modes that could enable the cavity photons to leak out to spatial infinity. Such BICs have recently been discovered in linear photonics \cite{hsu2016bound}. In the absence of any nonlinearity or interaction with the continuum, the bound cavity mode would be associated with the Hamiltonian  $\omega_a a^\dagger a$, with $\omega_a$ the linear cavity frequency and $a^{(\dagger)}$ the annihilation (creation) operator. We now take this system and introduce nonlinearity, as well as continuum couplings. The effect of nonlinearity is to add a term to the Hamiltonian $\beta\omega_a a^{\dagger 2}a^2$, with $\beta$ the dimensionless nonlinearity per photon. We define the Hamiltonian of the nonlinear cavity alone as $H_0/\hbar = \omega_a a^\dagger a + \beta\omega_a a^{\dagger 2}a^2$.

When such nonlinear cavities are coupled to a continuum of waveguide modes, a reservoir Hamiltonian is introduced, as well as its coupling to the cavity. In general, the coupling of a continuum mode at frequency $\omega$ to the cavity is frequency-dependent. In the frequency-independent-coupling case, a standard time-local Lindblad master equation can be derived. In our case however, it will be important to have the coupling be frequency-dependent so it can ``interact" with the photon-number-dependent resonance frequency $\omega_{n}-\omega_{n-1} = \omega_a + 2\beta\omega_a(n-1)$ to create a photon-number-dependent coupling to the reservoir (in other words, a photon-number-dependent loss). In this regime, a simple closed-form master equation is difficult to derive. In \cite{rivera2023creating}, an approximate master equation was derived, in a perturbative expansion around frequency-independent couplings. Using this approximate master equation, it was shown how Fock states could be stabilized. In what follows, we introduce a new effective model for nonlinear dissipation, which, while matching \cite{rivera2023creating} in the relevant limit, is essentially exact: holding for arbitrarily strong frequency-variations in the outcoupling, and also holding in the single-photon nonlinear regime which is of interest to us here. It is in this regime that the new results of this work show up.

The model is inspired by the textbook theory of dissipation in quantum optics: namely by considering dissipation as a beamsplitter which introduces an interaction with an auxiliary mode, and traces that mode out after the interaction \cite{mandel1995optical}. This model is illustrated in Fig \ref{fig1}B, and goes as follows: we pass an input light state $\rho_{\text{in}}$ through a sequence of nonlinear Mach-Zender interferometers (MZI; green box), each comprising a beamsplitter, Kerr nonlinearity, and another beamsplitter. The input mode and vacuum mode with respective annihilation operators $a,b$ are combined in the first beamsplitter with unitary operator $U_{bs}=e^{i\chi(a^\dagger b+ab^\dagger)}$. The Kerr nonlinearity involves the unitary evolution $e^{-iH_0\tau/\hbar}$ for time $\tau$, and the two modes are combined with another identical beamsplitter \footnote{It is worth noting that such nonlinear Mach-Zender interferometers were introduced in \cite{schmitt1998photon} to enable photon number squeezing.}. After each MZI, we trace out the output mode $b$ corresponding to the vacuum mode fed into the first beamsplitter. This model can be implemented numerically in a direct and straightforward way, and also readily enables consideration of ``driving'' by injecting non-vacuum states into the $b$ mode \cite{sloan2023driven}, as well as a treatment of nonlinear gain phenomena (similar to \cite{scully1999quantum,nguyen2023intense}, which would enable squeezing and other non-Gaussian quantum states at optical frequencies). It is however worth mentioning that in the limit of $\chi \ll 1$ (small splitting ratio; in the linear regime, small loss per beamsplitter), we can write an approximate ``update'' rule for the density matrix after one pass through a nonlinear MZI. That rule is (see Supplementary Information (SI) for derivation):
\begin{widetext}
\begin{equation}\label{eq:beamsplittereq}
\begin{gathered}
\rho_{n,n-k}\to e^{-i\omega_{n,n-k}\tau}\left(1-\chi^2\left(n+ne^{i\omega_{n,n-1}\tau}+(n-k)+(n-k)e^{-i\omega_{n-k,n-k-1}\tau}\right)\right)\rho_{n,n-k}\\
+e^{-i\omega_{n+1,n-k+1}\tau}\chi^2\sqrt{(n+1)(n-k+1)}
\left(1+e^{i\omega_{n+1,n}\tau}+e^{-i\omega_{n-k+1,n-k}\tau}+e^{i(\omega_{n+1,n}-\omega_{n-k+1,n-k})\tau}\right)\rho_{n+1,n-k+1},
\end{gathered}
\end{equation}
\end{widetext}
where $\rho_{n,n-k} = \langle n |\rho|n-k\rangle$. In the SI, we also show the equivalence to the approximate master equation of \cite{rivera2023creating} in the relevant limit. Before analyzing the consequences of this equation, we briefly mention that the ``chain-of-MZI'' model is essentially equivalent to an ``unfolded'' representation of the cavity in Fig. 1A: by considering each round trip-plus-nonlinear phase in the cavity as one of these MZIs. Therefore, the nonlinear interaction time $\tau$ in the MZI maps to $1/\gamma$ in the cavity of Fig. 1A. Meanwhile the parameter $\chi$ is related to $\kappa/\gamma$ (this mapping is rigorously shown in the SI). Now we utilize this model to study quantum state dynamics under this engineered dissipation.

To start, let us define the following function that organizes the physics: $K_1(\omega) \equiv \chi^2(1+e^{i\omega\tau})$. We refer to it as the ``loss function'' (see SI for justification). This is useful to define, as it can be seen that the discrete-time update equation of Eq. (1) is, in the limit $\chi \ll 1$, approximately equivalent to the following differential equation presented in \cite{rivera2023creating} for the density matrix (see SI for derivation): 
\begin{widetext}
\begin{eqnarray}\label{mastereq}
    \dot\rho_{n,n-k}=-i\omega_{n,n-k}\rho_{n,n-k}-\left(nK_1(\omega_{n,n-1})+(n-k)K_1^*(\omega_{n-k,n-k-1})\right)\rho_{n,n-k}\nonumber \\
+\sqrt{(n-k+1)(n+1)}\left(K_1(\omega_{n+1,n})+K_1^*(\omega_{n-k+1,n-k})\right)\rho_{n+1,n-k+1}.
\end{eqnarray}
\end{widetext}

\begin{figure}[t]
  \centering
  \includegraphics[width=3.25in]{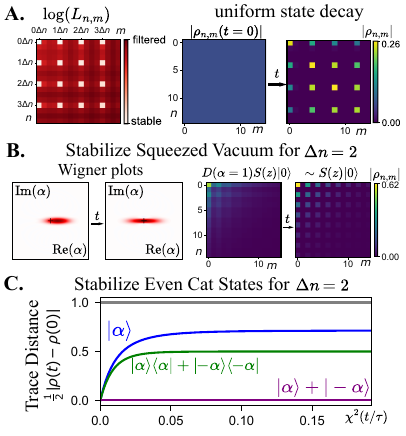}
\caption{\textbf{A.} Demonstration of periodically stable photon numbers and coherences. A phase state (uniform density matrix) decays into a coherent superposition of Fock states with photon numbers different by 4. \textbf{B.} By tuning the cavity, one can stabilize an arbitrary superposition of even photon numbers. For example, a displaced squeezed state decays to a squeezed vacuum with even photon parity. \textbf{C.} Dissipation of an initial coherent state, mixed superposition of coherent states, and cat state where even photon numbers are stabilized. The coherent and mixed states diverge from their initial states but the even cat state is stabilized.
}
\label{fig2}
\end{figure}

 We see that the loss function $K_1(\omega)$ vanishes periodically for certain frequencies ($\omega = (2m + 1)\pi/\tau$, with $m$ an integer); thus combining this with a nonlinear Kerr energy spectrum $\omega_{n,n-1}=\omega_a(1+2\beta(n-1))$ leads to the unique property of vanishing loss for periodic photon numbers. Furthermore, not only can we stabilize periodic photon numbers, certain coherences $\rho_{n,n-k}$ can be stabilized too. In Eq. (2) each density matrix entry $\rho_{n,m}$ decays with rate $L_{n,n-k}$ where 
\begin{equation}
\begin{gathered}
    L_{n,n-k}=\mathrm{Re}\left[nK_1(\omega_{n,n-1})+(n-k)K_1^*(\omega_{n-k,n-k-1})\right]\\
    =\chi^2n\left(1+\cos\left(\omega_a\tau(1+2\beta(n-1))\right)\right)\\
    +\chi^2(n-k)\left(1+\cos\left(\omega_a\tau(1+2\beta(n-k-1))\right)\right).
\end{gathered}
\end{equation}
We note that $L_{n,n-k}=0$ when $\mathrm{Re}(K_1(\omega_{n,n-1}))=\mathrm{Re}(K_1(\omega_{n-k,n-k-1}))=0$, so precisely setting $\omega_a\tau(1+2\beta (n_0-1))=\pi+2\pi m$ for integers $n_0,m$ can eliminate the cavity loss $L_{n_0,n_0}$ for photon number $n_0$. We also see that setting $\frac{\pi}{\omega_a\beta\tau}=\Delta n$ for an integer $\Delta n$ allows for the stabilization of photon numbers and coherences at multiples of $\Delta n$ away from $n_0$.
This property is illustrated in Fig \ref{fig2}A, showing the log-loss rates $\log(L_{n,m})$ corresponding to the cavity density matrix elements $\rho_{n,m}$. The photon numbers are stabilized at multiples of $\Delta n=4$ in this example, and we demonstrate the effect of this loss on an initialized  ``phase'' eigenstate $|\psi\rangle = (N+1)^{-1/2}\sum_{s=0}^N |s\rangle$\footnote{Here we set an arbitrary cutoff number for the phase state, but the behavior is the same for higher photon numbers.}. At long times, the phase state transforms into a coherent superposition of photon-number states spaced by four: a ``photon-number comb.'' 

If we change the cavity, for example by changing the free-spectral range in Fig \ref{fig1}A (left) and thus adjusting $\tau$, we can change the zeros of the loss to be ``every other photon number'' with $\Delta n=2$. In this case, important parity states such as squeezed states, even-order cat states, and approximate GKP states stay invariant under this dissipation. This is illustrated in Fig \ref{fig2}B, where we show how an initial displaced squeezed state exhibits long-time decay into a squeezed vacuum state $e^{\frac{1}{2}(z^*a^{2}-za^{\dagger 2})}|0\rangle$ with even photon parity. This is in contrast to linear loss, where the initial state evolves to the vacuum state. Furthermore, we demonstrate the stabilization of even photon parity cat states in Fig \ref{fig2}C. Starting from an initial coherent state $|\alpha\rangle$, a mixed coherent superposition $|\alpha\rangle\langle\alpha|+|-\alpha\rangle\langle-\alpha|$, and a cat state $|\alpha\rangle+|-\alpha\rangle$, we can find their deviations from the initial states throughout dissipative evolution. Here we find the distances from the initial states $T(\rho(t),\rho(0))=\frac12\mathrm{Tr}\left(\sqrt{(\rho(t)-\rho(0))^\dagger (\rho(t)-\rho(0))}\right)$ to evaluate how much the quantum states changed. We observe that coherent states and the $|\alpha\rangle\langle\alpha|+|-\alpha\rangle\langle-\alpha|$ states without even photon parity end up evolving into different quantum states through nonzero dissipation, but the even cat state $|\alpha\rangle+|-\alpha\rangle$ is stabilized. With the even cat state properties, this consistent with the expectation of zero loss of even photon numbers and dissipation of odd photon numbers.

\begin{figure*}[t]
    \centering
    \includegraphics[width=6.5in]{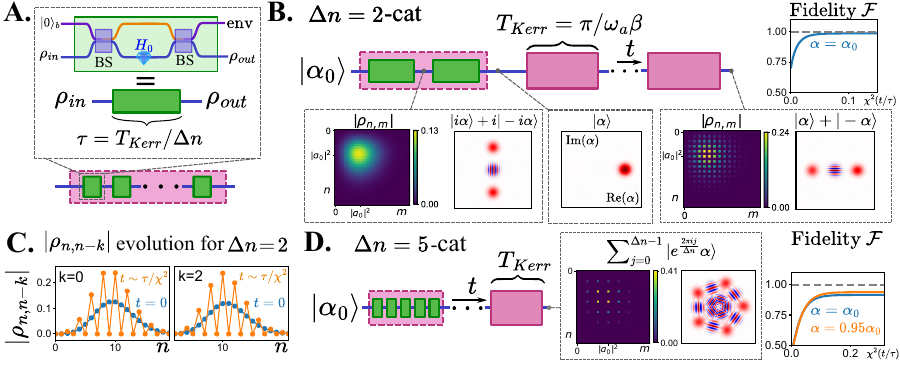}
    \caption{Generation of cat states through nonlinear loss.  \textbf{A.} Schematic of MZI units (green) of time step $\Delta t=1/\gamma$ making up a full Kerr period $T_{Kerr}=\pi/\omega_a\beta$. \textbf{B.} Sequence of MZI units for generating even cat states. For $T_{Kerr}\ll \tau/\chi^2$, the dynamics of purely Kerr unitary evolution and the production of intermediate $i$-cat states without parity properties are shown. This is in contrast with the final even cat state generation from free nonlinear loss with fidelity approaching $98.7\%$. Here, we use $\omega_a\beta\tau=\frac\pi2$, $\chi=0.01$, $\tau=200\pi/\omega_a$ and $\beta=2.5\times10^{-3}$. \textbf{C.} Photon populations and $k=2$ coherences in the time evolution of a coherent state in a $\Delta n=2$ photon comb. Not only do the even photon number populations increases, the $k=2$ coherences increase too. \textbf{D.} Generalization of $\Delta n$-legged cat state generation through free nonlinear loss. An example for $\Delta n=5$ is shown where the fidelities of generating a 5-legged $\alpha$ cat state with $\alpha=\alpha_0,\alpha=0.95\alpha_0$ approach $91.7\%,94.0\%$ respectively. Here, we have $\omega_a\beta\tau=\frac\pi5$, $\chi=0.003$, $\tau=201.4\pi/\omega_a$, $\beta\sim9.93\times10^{-4}$.}
    \label{fig3}
\end{figure*}

We can extend this beyond quantum state stabilization to achieve quantum state generation from classical states. In particular, we show how this dissipation creates large cat states from coherent inputs. In Fig \ref{fig3}A, we illustrate the units of the beamsplitter model that correspond to continuous nonlinear loss. We emphasize the period of free lossless Kerr evolution recoherences $T_{Kerr}=\frac{\pi}{\omega_a\beta}$ to distinguish Kerr unitary evolution against the dynamics of nonlinear loss. In Fig \ref{fig3}B (left gray box), we take a setup that stabilizes $\Delta n=2$-legged cat states, and illustrate the dynamics of \emph{pure Kerr nonlinear evolution} (no loss / no beamsplitters) for short term evolution (where $\chi^2 T_{osc}/\tau\ll1$) compared to the dynamics of \emph{nonlinear loss} (after time $t\sim \tau/\chi^2$ in the right gray box). Pure Kerr evolution conserves photon numbers, and therefore cannot produce canonical cat states of the form $|\psi\rangle \sim |\alpha\rangle \pm |-\alpha\rangle$, but rather produces what we call an ``$i$-cat state'' $|\psi\rangle \sim |i\alpha\rangle + i|-i\alpha\rangle$. Since this setup traps frequencies corresponding to even photon numbers through interference, the odd photon frequencies are maximally lossy and will be filtered out from the initial state. Through this natural photon parity selection, the cavity will dissipate into steady state solutions such as the even cat state (as shown in Fig \ref{fig2}C), allowing for cat state generation without control sequences.
We see that this scheme can generate cat states with high fidelity (top right plot), where the fidelities are calculated at integer time multiples of $T_{Kerr}$. Here, producing the cat state $|\psi\rangle\sim |\alpha\rangle+|-\alpha\rangle$ with $\alpha=\alpha_0$ from an initial $\alpha_0=\sqrt{10}$ coherent state approached $98.7\%$ fidelity. Due to the nonzero dissipation, and the fidelity towards an $\alpha=0.98\alpha_0$ cat state approached $\mathcal F=99.0\%$. Although the high cat state fidelities are not perfect, this scheme produces the important photon parity properties in the cat state generation, unlike the i-cat states produced through pure unitary evolution.

Generating a cat state $|\alpha\rangle+|-\alpha\rangle$ from an initial coherent state $|\alpha\rangle$ demands that both even photon probabilities and even coherences need to increase. In Fig \ref{fig3}C, we show the remarkable phenomenon where the $k=2$ coherences are not only stabilized, but increased during nonlinear dissipation. This property is not predicted by previous approximate models for dispersive nonlinear loss, e.g. in equation \ref{mastereq}, due to minor approximations in its derivation. The more exact beamsplitter model demonstrates this capability, as seen in the subtle differences in the two models' equations of motion as detailed in the SI. With more exact models, we can see how weakly non-Markovian can help engineer photon coherences in addition to probabilities.
Moreover, by controlling the geometry of the cavity, one can also stabilize $\frac{\pi}{\omega_a\beta\tau}=\Delta n$ parity photon numbers to generate $\Delta n$-legged cat states. We show an example for $\Delta n=5$ in Fig \ref{fig3}D. As before, free Kerr evolution can generate an $\Delta n$-legged ``$i$-cat state" after time $T_{Kerr}/\Delta n$, but the ``$i$-cat state" lacks the photon and coherence parities. After undergoing nonlinear dissipation, we can successfully filter out photon numbers without $\Delta n$ parity and generate an optical $\Delta n$-legged cat state $\sum_{j=0}^{\Delta n-1}|e^{\frac{2\pi i}{\Delta n}j}\alpha\rangle$. From an initial $\alpha_0=\sqrt{15}$ coherent state, we can generate an $\alpha=0.95\alpha_0$ 5-legged cat state with $94.0\%$ fidelity. Since the filtered photon number parity is adjustable from the cavity geometry, we are able to stabilize and generate arbitrary $\Delta n$-parity quantum states through free dissipation.

\section*{Conclusion}

Looking forward, we expect that these nonlinear dissipative dynamics can be fruitfully implemented in various systems where single-photon nonlinearities have been realized or are within reach: such systems include Rydberg-EIT systems and systems of exciton polaritons coupled to cavities with frequency-dependent outcouplers. Another interesting direction is to use strongly nonlinearities associated with high-cooperativity cavity QED systems. In such systems, when the number of polaritons is large, the Jaynes-Cummings dynamics act similarly to a Kerr nonlinearity. A major theoretical direction forward is designing new dissipators that deterministically generate other target quantum states from simple coherent inputs, e.g., GKP states. An interesting approach is to include second-order nonlinearities, which readily produce Gaussian states such as squeezed states. The combination of second- and third-order nonlinearities is expected to stabilize a much wider range of quantum states, due to the phase-sensitive nature of second-order nonlinearity. We conclude by mentioning that a major advantage of the nonlinear dissipators we put forth here is that, since they do not make use of pulse sequences, the bandwidth can be in principle very high, allowing future application of these principles to shape quantum states of \emph{ultrafast} light. For example, in a chain of nonlinear MZIs, provided that the outcoupler is broadband, the nonlinearity can act effectively even on ultrafast pulses of light, allowing the possibility of realizing non-Gaussian states of light in the picosecond and femtosecond regime.
\vspace{-0.2cm}
\section*{Acknowledgements}
\vspace{-0.2cm}
This material is based upon work sponsored in part by the U.S. Army DEVCOM ARL
Army Research Office through the MIT Institute for Soldier Nanotechnologies
under Cooperative Agreement number W911NF-23-2-0121. N.R. acknowledges the
support of a Junior Fellowship from the Harvard Society of
Fellows. J.S. acknowledges previous support
of a Mathworks Fellowship, as well as previous support from
a National Defense Science and Engineering Graduate (NDSEG) Fellowship (F-1730184536).
\bibliographystyle{unsrt}
\bibliography{apssamp}

\begin{thebibliography}{10}

\bibitem{gottesman2001encoding}
Daniel Gottesman, Alexei Kitaev, and John Preskill.
\newblock Encoding a qubit in an oscillator.
\newblock {\em Physical Review A}, 64(1):012310, 2001.

\bibitem{giovannetti2011advances}
Vittorio Giovannetti, Seth Lloyd, and Lorenzo Maccone.
\newblock Advances in quantum metrology.
\newblock {\em Nature photonics}, 5(4):222--229, 2011.

\bibitem{degen2017quantum}
Christian~L Degen, Friedemann Reinhard, and Paola Cappellaro.
\newblock Quantum sensing.
\newblock {\em Reviews of modern physics}, 89(3):035002, 2017.

\bibitem{wang2020efficient}
Christopher~S Wang, Jacob~C Curtis, Brian~J Lester, Yaxing Zhang, Yvonne~Y Gao,
  Jessica Freeze, Victor~S Batista, Patrick~H Vaccaro, Isaac~L Chuang, Luigi
  Frunzio, et~al.
\newblock Efficient multiphoton sampling of molecular vibronic spectra on a
  superconducting bosonic processor.
\newblock {\em Physical Review X}, 10(2):021060, 2020.

\bibitem{hofheinz2008generation}
Max Hofheinz, EM~Weig, M~Ansmann, Radoslaw~C Bialczak, Erik Lucero, M~Neeley,
  AD~O’connell, H~Wang, John~M Martinis, and AN~Cleland.
\newblock Generation of fock states in a superconducting quantum circuit.
\newblock {\em Nature}, 454(7202):310--314, 2008.

\bibitem{vlastakis2013deterministically}
Brian Vlastakis, Gerhard Kirchmair, Zaki Leghtas, Simon~E Nigg, Luigi Frunzio,
  Steven~M Girvin, Mazyar Mirrahimi, Michel~H Devoret, and Robert~J Schoelkopf.
\newblock Deterministically encoding quantum information using 100-photon
  schr{\"o}dinger cat states.
\newblock {\em Science}, 342(6158):607--610, 2013.

\bibitem{grimm2020stabilization}
Alexander Grimm, Nicholas~E Frattini, Shruti Puri, Shantanu~O Mundhada, Steven
  Touzard, Mazyar Mirrahimi, Steven~M Girvin, Shyam Shankar, and Michel~H
  Devoret.
\newblock Stabilization and operation of a kerr-cat qubit.
\newblock {\em Nature}, 584(7820):205--209, 2020.

\bibitem{campagne2020quantum}
Philippe Campagne-Ibarcq, Alec Eickbusch, Steven Touzard, Evan Zalys-Geller,
  Nicholas~E Frattini, Volodymyr~V Sivak, Philip Reinhold, Shruti Puri, Shyam
  Shankar, Robert~J Schoelkopf, et~al.
\newblock Quantum error correction of a qubit encoded in grid states of an
  oscillator.
\newblock {\em Nature}, 584(7821):368--372, 2020.

\bibitem{deng2023heisenberg}
Xiaowei Deng, Sai Li, Zi-Jie Chen, Zhongchu Ni, Yanyan Cai, Jiasheng Mai, Libo
  Zhang, Pan Zheng, Haifeng Yu, Chang-Ling Zou, et~al.
\newblock Heisenberg-limited quantum metrology using 100-photon fock states.
\newblock {\em arXiv preprint arXiv:2306.16919}, 2023.

\bibitem{peyronel2012quantum}
Thibault Peyronel, Ofer Firstenberg, Qi-Yu Liang, Sebastian Hofferberth,
  Alexey~V Gorshkov, Thomas Pohl, Mikhail~D Lukin, and Vladan Vuleti{\'c}.
\newblock Quantum nonlinear optics with single photons enabled by strongly
  interacting atoms.
\newblock {\em Nature}, 488(7409):57--60, 2012.

\bibitem{walther2006cavity}
Herbert Walther, Benjamin~TH Varcoe, Berthold-Georg Englert, and Thomas Becker.
\newblock Cavity quantum electrodynamics.
\newblock {\em Reports on Progress in Physics}, 69(5):1325, 2006.

\bibitem{delteil2019towards}
Aymeric Delteil, Thomas Fink, Anne Schade, Sven H{\"o}fling, Christian
  Schneider, and Ata{\c{c}} {\.I}mamo{\u{g}}lu.
\newblock Towards polariton blockade of confined exciton--polaritons.
\newblock {\em Nature materials}, 18(3):219--222, 2019.

\bibitem{munoz2019emergence}
Guillermo Mu{\~n}oz-Matutano, Andrew Wood, Mattias Johnsson, Xavier Vidal,
  Ben~Q Baragiola, Andreas Reinhard, Aristide Lema{\^\i}tre, Jacqueline Bloch,
  Alberto Amo, Gilles Nogues, et~al.
\newblock Emergence of quantum correlations from interacting fibre-cavity
  polaritons.
\newblock {\em Nature materials}, 18(3):213--218, 2019.

\bibitem{liran2023electrically}
Dror Liran, Jiaqi Hu, Nathanial Lydick, Hui Deng, Loren Pfeiffer, and Ronen
  Rapaport.
\newblock Electrically controlled dipolariton circuits.
\newblock {\em arXiv preprint arXiv:2308.08289}, 2023.

\bibitem{harrington2022engineered}
Patrick~M Harrington, Erich~J Mueller, and Kater~W Murch.
\newblock Engineered dissipation for quantum information science.
\newblock {\em Nature Reviews Physics}, 4(10):660--671, 2022.

\bibitem{hsu2016bound}
Chia~Wei Hsu, Bo~Zhen, A~Douglas Stone, John~D Joannopoulos, and Marin
  Solja{\v{c}}i{\'c}.
\newblock Bound states in the continuum.
\newblock {\em Nature Reviews Materials}, 1(9):1--13, 2016.

\bibitem{rivera2023creating}
Nicholas Rivera, Jamison Sloan, Yannick Salamin, John~D Joannopoulos, and Marin
  Solja{\v{c}}i{\'c}.
\newblock Creating large fock states and massively squeezed states in optics
  using systems with nonlinear bound states in the continuum.
\newblock {\em Proceedings of the National Academy of Sciences},
  120(9):e2219208120, 2023.

\bibitem{mandel1995optical}
Leonard Mandel and Emil Wolf.
\newblock {\em Optical coherence and quantum optics}.
\newblock Cambridge university press, 1995.

\bibitem{sloan2023driven}
Jamison Sloan, Nicholas Rivera, and Marin Solja{\v{c}}i{\'c}.
\newblock Driven-dissipative phases and dynamics in non-markovian nonlinear
  photonics.
\newblock {\em arXiv preprint arXiv:2309.09863}, 2023.

\bibitem{scully1999quantum}
Marlan~O Scully and M~Suhail Zubairy.
\newblock Quantum optics, 1999.

\bibitem{nguyen2023intense}
Linh Nguyen, Jamison Sloan, Nicholas Rivera, and Marin Solja{\v{c}}i{\'c}.
\newblock Intense squeezed light from lasers with sharply nonlinear gain at
  optical frequencies.
\newblock {\em Physical Review Letters}, 131(17):173801, 2023.

\bibitem{schmitt1998photon}
S~Schmitt, J~Ficker, M~Wolff, F~K{\"o}nig, A~Sizmann, and Gerd Leuchs.
\newblock Photon-number squeezed solitons from an asymmetric fiber-optic sagnac
  interferometer.
\newblock {\em Physical review letters}, 81(12):2446, 1998.

\end{thebibliography}

\end{document}